# Probing nanocrystalline grain dynamics in nanodevices

Sheng-Shiuan Yeh,[1] Wen-Yao Chang,[1] Juhn-Jong Lin [1,2,]*
[1]Institute of Physics, National Chiao Tung University, Hsinchu 300, Taiwan
[2]Department of Electrophysics, National Chiao Tung University, Hsinchu 300, Taiwan
*E-mail: jjlin@mail.nctu.edu.tw

## Abstract

Dynamical structural defects exist naturally in a wide variety of solids. They fluctuate temporally, and hence can deteriorate the performance of many electronic devices. Thus far, the entities of such dynamic objects have been identified to be individual atoms. On the other hand, it is a long-standing question whether a nanocrystalline grain constituted of a large number of atoms can switch, as a whole, reversibly like a dynamical atomic defect (i.e., a two-level system). This is an emergent issue considering the current development of nanodevices with ultralow electrical noise, qubits with long quantum coherence time, and nanoelectromechanical system (NEMS) sensors with ultrahigh resolution. Here we demonstrate experimental observations of dynamic nanocrystalline grains which repeatedly switch between two or more metastable coordinate states. We study temporal resistance fluctuations in thin ruthenium dioxide ($RuO_2$) metal nanowires and extract microscopic parameters including relaxation time scales, mobile grain sizes, and the bonding strengths of nanograin boundaries. Such material parameters are not obtainable by other experimental approaches. When combined with previous *in-situ* high-resolution transmission electron microscopy (HRTEM), our electrical method can be used to infer rich information about the structural dynamics of a wide variety of nanodevices and new 2D materials.

## Introduction

Dynamical structural defects occur commonly in artificially grown low-dimensional materials. Such unwanted mobile objects (also termed "fluctuators" or moving scattering centers) have been identified as dynamical atomic defects (*1-3*), which repeatedly switch back and forth between two nearby lattice positions with similar configurational energies. Theoretically, such fluctuators are modeled as atomic two-level systems (TLSs) (*4*, *5*). In polycrystalline materials, as the grain size is reduced from the millimeter to submicron scale, the yield stress, flow stress and hardness *increase*, known as the Hall-Petch effect (*6*). Nevertheless, if the grain size is further reduced to the nanometer scale, the flow stress will *decrease* with decreasing grain size, resulting in the reverse Hall-Petch effect (*7-10*). This reverse effect stems from the fact that, for a nanocrystalline grain, a large fraction of its constituent atoms reside at or nearby grain boundaries, which are elastically softer than the bulk lattice atoms (*7*). Thus, plastic deformation occurs more easily for a polycrystalline solid containing nanometer-sized grains. Microscopically, a soft grain boundary reflects a weak bonding strength between grain-boundary atoms (*11*, *12*). Previously, it was found in both *in-situ* HRTEM and molecular dynamics simulation studies of polycrystalline metal films that nanograin rotations (*8*, *13*, *14*) and boundary sliding (*6*, *9*) can occur in specimens *under stress*. The fundamentally and technically important question of whether a nanograin with soft grain boundaries can switch reversibly between two (or a few) nearby structural configurations in the absence of external stress then follows. In this case, we may model a fluctuating nanocrystalline grain by a "granular TLS," as schematically depicted in Figs. 1A and 1B. It was recently found that domain boundaries could be induced in a magnetic skyrmion lattice (*15*) by the $FeGe_{1-x}Si_x$ crystalline grain boundaries (*16*). The switches of a crystalline grain, causing domain boundary fluctuations, may thus modify the dynamic properties of magnetic skyrmions. In industrial applications, the requirement for mass production and integrated-circuit scaling results in the polycrystalline features of practical devices. For example, large-area graphene films produced by chemical vapor deposition are naturally polycrystalline (*17*, *18*). Other examples include thin-film polycrystalline silicon nanowire biosensors (*19*) and wafer-scale reduced polycrystalline graphene oxide films for NEMS devices (*20*). Thus, it is highly desirable to study whether a nanocrystalline grain can be dynamic and act as a granular TLS. As mentioned above, the presence of atomic and granular TLSs will deteriorate the performance of nanoelectronic devices (*21*), hinder the quantum state coherence time in qubits (*22*), and limit the quality factors of NEMS sensors (*23*).

In this study, we demonstrate that a nanograin can spontaneously and reversibly fluctuate in the absence of externally applied stress. We use thin $RuO_2$ nanowires grown via two different methods to address this problem. Nanowires A–C were grown via metal-organic chemical vapor deposition (MOCVD), whereas nanowire D was grown via thermal evaporation (Materials and Methods). The advantages of using $RuO_2$ nanowires are as follows. (i) Our $RuO_2$ nanowires are polycrystalline, containing nanometer-sized grains. Figure 2A shows a HRTEM image of a ~ 7 nm diameter $RuO_2$ nanowire from the same batch as nanowires A–C. Close inspection of the image indicates the presence of several nanocrystalline grains with small diameters of 1–6 nm, labeled $G_1$–$G_6$ (fig. S1 shows more HRTEM images). The formation of nanograins probably results from a large strain induced by the radial confinement in the nanowire geometry (*24*). Figure 2B shows two HRTEM images of a ~ 30 nm (left) and a ~ 90 nm (right) diameter $RuO_2$ nanowires from the same batch as nanowire D. The nanowires exhibit crystallites (nanograins) of tens nm in size. Stacking faults, indicated by arrows, along the axial direction are observed, which can result in weak bonding and induce grain (crystallite) boundary sliding (*25*). Previously, we have observed profound universal conductance fluctuations (UCFs) in single $RuO_2$ nanowires at low temperatures, suggesting the existence of a large number of vibrant (atomic) TLSs (*26*). A significant portion of the atomic TLSs should reside at or nearby grain boundaries (*11*, *12*), which in turn may lead to weak nanograin bonding. Thus, the existence of granular TLSs can be expected. (ii) $RuO_2$ nanowires are metallic (*27*, *28*), allowing us to employ a high-resolution electrical method to detect nanograin motion. In contrast to the case of doped semiconductors, where temporal resistance fluctuations originate from both mobility and carrier number fluctuations (*29*), the resistance fluctuations in $RuO_2$ nanowires occur solely due to mobility fluctuations arising from electron scattering off moving defects. The temperature dependence of the averaged resistivity $\langle \rho(T) \rangle$, for our nanowires below 300 K obeys the Boltzmann transport equation, ignoring the quantum-interference corrections at liquid-helium temperatures (*28*). (iii) $RuO_2$ nanowires are thermally and chemically stable under ambient conditions (*27*). Thus, their electrical properties are stable and do not deteriorate under repeated resistance measurements. In short, thin $RuO_2$ metal nanowires are ideal candidates for in-depth studies of nanograin dynamics.

We measure the temporal resistivity fluctuations, $\rho(t)$, of single $RuO_2$ nanowires at room temperature. Owing to the large size of a granular TLS compared to that of an atomic TLS, the repeated switches of a granular TLS should result in marked random telegraph noise (RTN). The RTN can be heuristically characterized by two (or more) histogram peaks, as schematically depicted in Fig. 1C. The separation between the two histogram peaks arises due to the motion of a granular TLS, which causes a notable resistivity jump, hereafter denoted by $\Delta\rho_{grain}$. This is a type of *collective* motion of all the atoms in the grain. The finite width of each histogram peak represents the standard deviation for each resistivity state, hereafter denoted by $\Delta\rho_{sd}$, which is caused by the background noise plus a contribution due to the *independent* fluctuations of a large number of *uncorrelated* atomic TLSs coexisting in the nanowire. Thus far, the RTN due to granular TLSs has been reported only rarely, while the RTN due to atomic TLSs was observed in, *e.g.*, metal nanoconstrictions (*2, 30*). Unlike the *in-situ* HRTEM studies which can only examine a series of static images (for example, with frame rate of 1/30 sec (*14*)) for a certain nanograin configuration, our electrical method allows us to explore the dynamic processes, relaxation times, and geometrical sizes of granular TLSs. Consequently, the mechanical bonding strengths of switching granular TLSs can be inferred. Moreover, we emphasize that our method is noninvasive, while the HRTEM technique often employs high-energy electron beams of ≈ 100–200 keV, which readily introduce artifacts, such as dislocations and new defects, in the specimen (*31, 32*). For those granular TLSs which are embedded inside the bulk of a nanodevice and difficult to detect by, for instance, the HRTEM, our electrical method is well applicable. Figure 2C shows an SEM image for nanowire A together with our electrical measurement circuit (Materials and Methods). In this study, the $\rho(t)$ values of nanowires A–C (nanowire D) were registered per $\simeq 1$ ms for 30 s (8 s). Our nanowire parameters are listed in Table 1.

We reiterate that our electrical method can be applied to study a wide variety of low-dimensional materials which are highly topical. The presence of grain boundaries in large-area graphene (*17, 18, 33*) and 2D material films will not only cause large electrical noise (*34*) but generate granular TLSs which are yet to be explored (*17, 35*).

## Results

### Theory

**1/*f* noise and Hooge parameter.** The atomic TLS density, $n_{TLS}$, in a specimen can be evaluated through low-frequency noise studies. For an ohmic conductor with a relatively flat distribution of activation energy for the atomic TLSs, the resistivity noise power spectrum density (PSD) exhibits a $1/f^{\alpha}$ frequency dependence, with the exponent $\alpha \approx 1$ (*36, 37*). By applying a current, the resistivity noise is transformed to voltage noise with the PSD described by (*38*)

$$S_V = \frac{\gamma}{N_c}\frac{\langle V\rangle^2}{f^{\alpha}} + S_V^0 , \qquad (1)$$

where the Hooge parameter $\gamma$ characterizes the normalized noise amplitude, $N_c = n_e \times \mathrm{vol}$ is the total conduction electron number in the sample ($n_e$ being the conduction electron density, and vol being the sample volume), $V$ is the applied voltage drop across the sample, $\langle \cdots \rangle$ denotes the time average, and $S_V^0$ is the background noise PSD including the Johnson-Nyquist noise and the circuit noise. The Hooge parameter is theoretically expressed by (*39*): $\gamma \approx \left(n_e l \langle \sigma \rangle\right)^2 \left(n_{TLS}/n_e\right) / \ln\left(\omega_{\max}/\omega_{\min}\right)$, where $l = v_F \tau_{tot}$ is the electron mean free path, $v_F$ is the Fermi velocity, $1/\tau_{tot} = 1/\tau_e + 1/\tau_{in}$ is the total electron scattering rate ($\tau_e$ ($\tau_{in}$) being the elastic (inelastic) scattering time), $\langle \sigma \rangle$ is the averaged electron scattering cross-section, and $\omega_{\max}$ and $\omega_{\min}$ are respectively the upper- and lower-bound cutoff frequencies for the $1/f$ noise. For typical metals at 300 K, it was previously found that $\ln\left(\omega_{\max}/\omega_{\min}\right) \approx 5-25$ (*37, 39*). Hence, we take $\ln\left(\omega_{\max}/\omega_{\min}\right) \simeq 10$ in the following analysis. (This approximation may lead to an uncertainty in the extracted $n_{TLS}$ value by a factor of ~ 2.) This expression for $\gamma$ can be rewritten into the following form to illustrate the linear dependence of $\gamma$ on $n_{TLS}$:

$$\gamma = \frac{e^{-4}}{\ln\left(\omega_{\max}/\omega_{\min}\right)}\left(n_{TLS}\frac{m^2}{n_e}\frac{v_F^{\,2}}{\rho^2}\langle \sigma \rangle^2\right) , \qquad (2)$$

where $e$ is the electronic charge, and $m$ is the effective electron mass. In applying Eq. (2), we may take $\langle\sigma\rangle \approx 4\pi/k_F^{\ 2}$ (*39-41*), where $k_F$ is the Fermi wavenumber. Therefore, the value of $n_{TLS}$ can be directly calculated from the measured $\gamma$ value. In $RuO_2$ nanowires (*42, 43*), we have the relevant parameters $m = 1.4 m_e$ ($m_e$ being the free electron mass), $k_F \approx 1\times10^{10}$ m$^{-1}$, $v_F \approx 8.2\times10^5$ m/s, and $n_e \approx 5\times10^{28}$ m$^{-3}$. Then, the Hooge parameter is $\gamma \approx 5.3\times10^{-41}\left(n_{TLS}/\rho^2\right)$, where $n_{TLS}$ is in m$^{-3}$, and $\rho$ is in Ω m. Owing to the uncertainties in the evaluation of electronic parameters, our estimates of $\gamma$ and $n_{TLS}$ are expected to be correct to within a factor of ~ 2–3.

**Resistivity variation due to one atomic TLS.** We consider the electron scattering off atomic TLSs at a microscopic level. For an atomic TLS, the electron scattering cross-section $\sigma_i$ differs for the two metastable coordinate states by an amount of $\delta\sigma_i$, where $i$ denotes the $i$th atomic TLS. The variation in the scattering cross-section leads to a resistivity change given by (*41*),

$$\delta\rho_i = \frac{m v_F}{n_e e^2}\frac{\delta\sigma_i}{\text{vol}} . \qquad (3)$$

In general, the magnitude of $\delta\sigma_i$ differs for different atomic TLSs. However, as discussed, we may assume $\langle\delta\sigma_i\rangle \approx \langle\sigma\rangle \approx 4\pi / k_F^{\ 2}$ in the following analyses (*39-41*). Based on Eq. (3), we show below that the total number of atomic TLSs in a nanodevice, as well as the geometrical size of a granular TLS, can be inferred from the measured $\rho(t)$.

**Resistivity variation due to a large number of independent atomic TLSs.** For many *uncorrelated* atomic TLSs switching at random, the root-mean-square (rms) value of the resultant variation of resistivity, $\Delta\rho_{TLS}$, can be calculated in terms of the well-established local-interference model (*39, 41, 44*) which is valid for $\tau_{in} \ll \tau_e$ (this criterion is pertinent to our measurements carried out at room temperature). Here the electron phase memory is lost between successive elastic scattering events and no phase-coherence effect is of concern (*45*). Under such conditions, Pelz and Clarke have formulated a semi-classical expression for the rms resistivity variation given by (*39, 41*):

$$\Delta\rho_{TLS} \approx \sqrt{N_{TLS}} \times \Delta\rho_i \approx \frac{m v_F}{n_e e^2}\frac{\sqrt{N_{TLS}}}{\text{vol}}\langle\sigma\rangle , \qquad (4)$$

where $N_{TLS} = n_{TLS} \times \text{vol}$ is the total number of atomic TLSs in the nanostructure, and $\Delta\rho_i$ denotes the rms value of $\delta\rho_i$.

**Resistivity jump due to a granular TLS.** We propose that the geometrical size of a nanoscale granular TLS can be estimated from the measured RTN as follows. We treat the motion of a nanograin composed of $N_{grain}$ atoms as a *collective* motion of all the atoms in the grain. The resultant variation of resistivity $\Delta\rho_{grain}$ may then be approximated by

$$\Delta\rho_{grain} \approx N_{grain} \times \Delta\rho_i \approx \frac{mv_F}{n_e e^2} \frac{N_{grain}}{\text{vol}} \langle \sigma \rangle \quad . \qquad (5)$$

Note that we presume the resultant resistivity variation scale with the number of effective scattering centers $N_{grain}$, because now we deal with a kind of collective motion. In contrast, for uncorrelated motions of $N_{TLS}$ atomic TLSs, the resultant resistivity variation scales with $\sqrt{N_{TLS}}$, as given in Eq. (4). Thus, the $N_{grain}$ value of a mobile nanograin can be calculated from Eq. (5), using the measured $\Delta\rho_{grain}$ value, which is the separation between the neighboring histogram peaks illustrated in Fig. 1C. We shall see that the $N_{grain}$ values thus inferred render mobile nanograin sizes in good accord with those detected by the HRTEM. Thus, Eq. (5) can be considered a satisfactory approximation to evaluate $N_{grain}$ values. The amplitude of $\Delta\rho_{grain}$ varies inversely with the sample volume $\text{vol}$, and hence the effect of nanograin motions on the sample resistivity can only be seen in small specimens.

## Experimental Results

**Extracting atomic TLS density from 1/*f* noise measurements.** Figures 3A–D show the measured $S_V$ for nanowires A–D at 300 K, and Fig. 3E shows the bias voltage dependence of $S_V$ for nanowires A–C, as indicated. We found $S_V \propto 1/f^{\alpha}$, with $\alpha \approx 0.96 - 1.23$, as well as $S_V \propto \langle V \rangle^2$ for nanowires A–C, as predicted by Eq. (1). These results suggest that our measured $1/f$ noise must originate from the resistivity fluctuations due to a large number of atomic TLSs. From the linear fits (straight lines) in Fig. 3E, we obtain the Hooge parameter $\gamma \approx 5\times10^{-3}$ (see Table 1). This magnitude of $\gamma$ is similar to those found in typical polycrystalline metal films (*46*). From Eq. (2), we calculate the atomic TLS density to be $n_{TLS} \approx 3\times10^{26}$ m$^{-3}$ in nanowires A–C. Theoretically, defects (scattering centers) in a specimen can be categorized into two groups: static defects and dynamic defects (namely, the atomic TLSs). The total resistivity $\rho$ is then treated as the sum of the resistivity of individual conduction electrons scattering off each defect (*41*), i.e., $\rho = \sum_i \rho_i \simeq \frac{mv_F}{n_e e^2}\frac{N_{defect}}{\mathrm{vol}}\langle\sigma\rangle = \frac{mv_F}{n_e e^2} n_{defect}\langle\sigma\rangle$, where $\rho_i$ is the resistivity due to the scattering off the *i*th defect, and $N_{defect}$ is the total number of defects in the nanowire. The total defect density $n_{defect} = N_{defect}/\mathrm{vol}$ can be directly calculated from the measured resistivity. We obtain a value of $n_{defect} \approx 1.8\times10^{28}$ m$^{-3}$ for nanowires A–C. Therefore, the TLSs amount to $\approx 0.3\%$ of the number of lattice atoms, or $\approx 2\%$ of the number of lattice defects (scattering centers) in the nanowire. Alternatively, the total number of atomic TLSs in nanowires A–C is on the order of $N_{TLS} = n_{TLS}\times\mathrm{vol} \approx (1-2)\times10^5$. Although our inferred $n_{defect}$ value seems to be somewhat high, it is in line with the observed small relative resistivity ratio (RRR) of $\rho(300\ \mathrm{K})/\rho(4\ \mathrm{K}) \approx 1.5-2$ in our nanowires (*28*), compared with that (RRR ≈ 100) in bulk $RuO_2$ single crystals (*47*). This defect density gives rise to an electron mean free path of ~ 0.4–0.5 nm, about twice the interatomic spacing. The apparent overestimate of $n_{defect}$ may partly arise from the uncertainties in the electronic parameters.

The atomic TLSs can originate from oxygen vacancies. As a control test, we have studied the $1/f$ noise of a few sputtered $RuO_2$ films that underwent thermal annealing in oxygen and argon. We found that those films annealed in oxygen possessed lower $1/f$ noise PSD, demonstrating an important role of oxygen vacancies. In addition, because a $RuO_2$ surface is readily hydrogenated in the atmosphere (*48*), hydrogen atoms adsorbed by oxygen atoms at the nanowire surface may possibly form atomic TLSs (*49*).

In sharp contrast, our measured $S_V$ in nanowire D reveals a distinct Lorentzian, instead of inverse frequency, dependence (Fig. 3D), suggesting that the noise behavior is dominated by the (slow) fluctuations of a single (large) TLS. In the following discussion, we shall demonstrate that this Lorentzian behavior originate from the motion of a single nanometer-sized granular TLS.

**Histogram plot of temporal resistivity fluctuations.** We demonstrate that the dynamics of granular TLSs can be examined in detail by analyzing the temporal resistivity fluctuations $\rho(t)$.

Figure 4A shows $\rho(t)$ for nanowire A, obtained over 30 s from the measured $V(t)$ with a small current of $I \simeq 0.5$ μA. This applied current corresponds to one hundredth of the breakdown current density for $RuO_2$ nanowires (*50*), and the current–induced electromigration was confirmed to be absent in our measurements (*51*). Such a small applied current also will not cause any diffusion defects along grain boundaries (*52*). In order to illustrate the details, we plot a small section of the $\rho(t)$ data in Figs. 4B–D for nanowires A–C, respectively. Close inspection reveals that each nanowire exhibits a number of preferred resistivity values (states), as indicated by the horizontal red dashed lines. The nanowire resistivity reversibly switches between these preferred states. Figures 4F–H show the corresponding histogram plots for $\rho(t)$ recorded over 30 s for these three nanowires, as indicated. We see that there exist four observable histogram peaks for nanowire A (labeled by $\rho_{A1} - \rho_{A4}$) and nanowire B (labeled by $\rho_{B1} - \rho_{B4}$), and eight observable histogram peaks for nanowire C (labeled by $\rho_{C1} - \rho_{C8}$). For example, the four histogram peaks from $\rho_{A1}$ to $\rho_{A4}$ for nanowire A correspond to the four red dashed lines in Figs. 4A and 4B. In each nanowire, while there exist a few preferred resistivity values (states), one of them clearly predominates, with a corresponding histogram peak much higher than the others. Such predominance implies that the nanowire visits this particular nanograin configuration much more frequently than it visits other nanograin configurations. In other words, the nanowire stays significantly longer in this metastable structural configuration than it stays in others. We found that nanowires A, B and C spent approximately 56%, 69% and 36% of the measurement time in their respective predominant histogram peaks.

The situation in nanowire D is distinctly different. Figure 4J shows $\rho(t)$ recorded over 8 s for this nanowire, with the corresponding histogram plotted in Fig. 4K. Two metastable resistivity values are clearly observed. Moreover, the two corresponding histogram peaks are of equal height, with each peak being well described by a Gaussian distribution (red curves, section S1).

For comparison, Fig. 4E shows $\rho(t)$ for an ordinary 10 kΩ metal-film resistor, with the corresponding histogram plotted in Fig. 4I. Clearly, there exists only one preferred resistance state. Due to its macroscopic size (≈ 0.2 $mm^3$), the motions of granular TLSs, if any exist, will not cause an observable resistance jump, because Eq. (5) predicts an inverse sample volume dependence (see section S2 for Gaussian functions and fig. S2 for $1/f$ noise PSD). Note that the measured resistance fluctuations are now due to the sample thermal noise together with the background noise from our electronics. The relative fluctuation amplitudes are 2–3 orders of magnitudes lower than those shown in Figs. 4A–D. Therefore, the high noise levels revealed in Figs. 4A–D must originate from nanowires A–C and not the experimental setup.

**Estimating the geometrical size of a mobile granular TLS.** In Figs. 4F-H, the typical separation between neighboring histogram peaks, $\Delta\rho_{grain}$, is significantly larger than the resistivity variation that would result from the switch of an atomic TLS. The observed separations between neighboring histogram peaks are $\Delta\rho_{grain} \approx 0.07$, $\approx 0.06$ and $\approx 0.02$ μΩ cm in nanowires A–C, respectively. Numerically, the prediction of Eq. (3) for the resistivity variation due to the switch of a single atomic TLS is on the order of $\delta\rho_i \approx (1-2)\times 10^{-5}$ μΩ cm in nanowires A–C. Figures 4F-H further show that, for a given nanowire, all the histogram peaks can be described by a set of Gaussian functions (blue curves, table S1) with standard deviations similar to within $\pm 20\%$. The sum (red curve) of all the Gaussian functions then well describes the entire histogram plot. Thus, the full histogram can be explained as originating from the coexistence of a large number of atomic TLSs and a single (few) granular TLS(s) in the nanowire. While the uncorrelated fluctuations of individual atomic TLSs together with the background noise from our experimental setup lead to a finite Gaussian width for each histogram peak, the discrete motion of a granular TLS leads to a jump of the preferred resistivity value (state). The width for all the histogram peaks should be similar, because the measurement conditions were fixed. For nanowires A and B, we can estimate from Eq. (4) that the independent fluctuations of the uncorrelated atomic TLSs contribute a resistivity variation of $\Delta\rho_{TLS} \approx \sqrt{N_{TLS}} \times \Delta\rho_i \approx 0.009$ μΩ cm, i.e., ~ 30% of the width of the histogram peak. Similarly, we estimate $\Delta\rho_{TLS} \approx 0.006$ μΩ cm, or about 50% of the peak width, for nanowire C. We recall that the full width at half maximum (FWHM) of a histogram peak is mathematically given by $\mathrm{FWHM} \simeq 2.35\Delta\rho_{sd}$.

Conceptually, the multiple histogram peaks can be envisioned as stemming from a mobile nanograin that repeatedly switches back and forth between a few metastable configurations (coordinate states). Indeed, recent *in-situ* HRTEM (*13, 14*) as well as theoretical molecular dynamics simulation studies (*8*) have unambiguously discovered the existence of mobile nanocrystalline grains. In particular, the HRTEM studies found that dislocations at grain boundaries can cause strain fields, leading to grain rotations (*13*). These independent structural TEM and theoretical simulation studies provide direct support for our electrical results. That is, a nanograin can move and that the resistance jumps we observe can reasonably be ascribed to the motions of granular TLSs. Furthermore, Figs. 4F-H indicate that each nanograin switch causes a similar $\Delta\rho_{grain}$ in a given nanowire. This significant observation suggests that the change of preferred resistivity states must be caused by the same granular TLS, invoking a fixed number of $N_{grain}$ in Eq. (5). In other words, we have detected a predominant mobile nanograin in each nanowire under our measurement conditions (see further discussion below).

The situation in nanowire D is distinct. Figure 4J implies that the predominant mobile grain possesses two metastable configurations and a switch between the two configurations causes a large resistivity jump of $\Delta\rho_{grain} \approx 1$ μΩ cm, corresponding to a *giant* relative resistivity change of $\Delta\rho_{grain} / \rho \simeq 1.4\%$. Indeed, the marked bi-state fluctuations are responsible for the Lorentzian noise PSD observed in Fig. 3D. For comparison, we found a much smaller ratio on the order of $\Delta\rho_{grain} / \rho \approx (1-4)\times 10^{-4}$ in nanowires A–C.

Substituting the measured $\Delta\rho_{grain}$ values into Eq. (5), we obtain $N_{grain} \approx 2.7\times 10^{3}$, $\approx 2.6\times 10^{3}$, $\approx 1.6\times 10^{3}$ and $\approx 3.0\times 10^{6}$ in nanowires A–D, respectively. Assuming spherical shape, we estimate from $N_{grain}$ the mobile grain size (diameter) to be ≈ 3.7, ≈ 3.7, ≈ 3.1, and ≈ 39 nm in nanowires A–D, respectively. Note that the mobile grain in nanowire D is one order of magnitude larger than the other nanowires. The formation of a large nanocrystalline grain probably results from a relatively weak radial stress embedded in this comparatively large diameter nanowire (*24*). Besides, the grain size may be affected by a different growth method. In any case, the giant resistance jumps in nanowire D can be ascribed to reversible motions of a large granular TLS, i.e., a large cluster of atoms, as has long been proposed and speculated in literature (*4, 53*).

In 1990, Giordano and Schuler (*53*) observed giant resistance fluctuations ( $\Delta R/R \approx 0.2\%$ ) in a 17 nm diameter Pb-In nanowire from electrical-transport measurements at ~ 100 K. They tentatively attributed their results to reversible motions of objects composed of large clusters of defects. At the time, it was not clear whether a large object (a group of atoms) can move reversibly. Their giant resistance fluctuations can now be understood and identified with the resistivity variation given by Eq. (5). Indeed, a simple estimate gives a value of $N_{grain} \sim 10^{5}$ atoms in their mobile grain, corresponding to a geometrical size of ~ 20 nm, in line with our result for nanowire D. Such good consistency between two independent studies further justifies the validity of Eq. (5). While Giordano and Schuler (*53*) only observed giant resistance fluctuations in Pb-In nanowires smaller than 20 nm in diameter, we have observed giant resistance fluctuations even in a 90 nm diameter $RuO_2$ nanowire, confirming that TLSs occur commonly in $RuO_2$ nanostructures (*26*). We have previously studied the electrical transport properties of several metallic nanowires, such as indium tin oxide (ITO) nanowires (*54*) and In-doped ZnO nanowires (*55*), but did not observe resistance fluctuations $\rho(t)$ as notable as those inherent in $RuO_2$ nanowires.

It is instructive to distinguish why the noise PSD obeys a Lorentzian dependence in nanowire D, while it obeys a $1/f$ dependence in nanowires A–C, even though grain motions occur in all nanowires. The frequency dependence of the noise PSD is mainly determined by two factors. The first factor is the relative size of $\Delta\rho_{grain}$ to the width of the histogram peak $2\Delta\rho_{sd}$. We found the ratios are $\Delta\rho_{grain}/(2\Delta\rho_{sd}) \approx 2.3$, ≈ 2.1, and ≈ 2.0 in nanowires A–C, respectively. The second factor is the relative height among the histogram peaks characterizing the different preferred resistivity states. Figures 4F–H show that the height of the predominant histogram peak is several times of those of the rest peaks in each nanowire. This highest histogram peak thus largely governs the measured $\rho(t)$ behavior, because the mobile nanograin spends most of the time in this particular metastable configuration, as discussed. Under such conditions, the PSD is mainly caused by the independent fluctuations of the $N_{TLS}$ individual atomic TLSs, giving rise to the usual $1/f$ feature. On the other hand, in nanowire D, we observed a much larger ratio $\Delta\rho_{grain}/(2\Delta\rho_{sd}) \approx 4.7$ with only two histogram peaks of equal height, suggesting that the large mobile nanocrystalline grain spent nearly the same time at the two equivalent metastable configurations. These repeated bi-state fluctuations governed the noise PSD, leading to a Lorentzian behavior found in Fig. 3D (*56*).

There exist fast fluctuations in the high and low resistivity states of nanowire D, Fig. 4J, which should cause 1/*f* noise. However, under the small bias voltage (64 μV) applied in Fig. 3D, we expect a 1/*f* noise PSD of $S_V < 10^{-19}$ $V^2$/Hz at 1 Hz, which is negligible compared to the measured Lorentzian (together with the background white) noise level.

**Extracting the relaxation time of a mobile granular TLS.** Having demonstrated that a nanocrystalline grain can move, we turn to evaluating the relaxation time, $\tau_0$, and potential barrier height, $V_B$, of a mobile granular TLS, Fig. 1B. For nanowires A–C, the average relaxation time for each metastable nanograin configuration can be extracted from $\rho(t)$. For a given preferred resistivity value, and assuming the relaxation-time approximation, the probability for the nanowire resistivity to remain at this value at time $t > 0$ is (*57*): $P(t) = P(0)\exp(-t/\tau_0)$, where $P(0)$ is the probability for taking this particular state at $t = 0$. As $t$ increases, a granular TLS may suddenly fluctuate, causing the nanowire to switch away from this state. Figures 5A–C plot the variation of the number of events (counts) with dwell time $t_d$ for each preferred resistivity value (state) observed in nanowires A–C, as indicated (section S3). Clearly, we find exponential time dependence in all cases, implying that a constant escape rate can be defined for the responsible metastable nanograin configuration. Furthermore, this exponential dependence suggests that, regardless of its fluctuating history, once the nanograin switches into a certain metastable configuration, the subsequent escape process from this configuration takes the same relaxation rate $1/\tau_0$. Microscopically, a constant escape rate implies a time-independent $V_B$ for the metastable nanograin configuration. A constant $V_B$ further implies that the responsible granular TLS must fluctuate around a fixed position, instead of moving away. That is, our low-current electrical measurements are noninvasive. From the linear fits (straight lines) in Figs. 5A–C, we obtain the relaxation time to lie in the range $\tau_0 \sim 0.4-6$ ms for all the metastable nanograin configurations detected in nanowires A–C. The extracted $\tau_0$ value for each configuration is listed in Figs. 5A–C in the parentheses. For a given nanowire, the extracted $\tau_0$ is longest for the highest histogram peak, because the nanowire spends most of its time in this particular resistivity state. In other words, once the mobile nanograin switches into this preferred metastable state, it is comparatively difficult for it to switch out. In fact, the $\tau_0$ value decreases monotonically with histogram peak height.

For nanowire D, due to a long sub-second dwell time period as can be directly seen in Fig. 4J, the number of switching events between the high and low resistivity states during our measurement time period is too few to allow a meaningful statistical analysis. Instead, the relaxation time for the metastable configuration can be inferred from the Lorentzian PSD behavior. The voltage noise PSD caused by one TLS is given by (*56*)

$$S_V = \frac{A\tau}{1 + 4\pi^2 f^2 \tau^2} + S_V^0 \quad , \qquad (6)$$

where $A$ is a constant, $1/\tau = 1/\tau_{0,1} + 1/\tau_{0,2}$, with $\tau_{0,1(2)}$ being the relaxation time for the high and low resistivity states, and $S_V^0$ is the background noise PSD. Our observation of two equivalent histogram peaks in Fig. 4K suggests that $\tau_{0,1} \simeq \tau_{0,2}$. Thus, from a least-squares fit of the data in Fig. 3D to the prediction of Eq. (6), we obtain $\tau_{0,1(2)} \approx 2\tau \approx 0.20$ s. Note that this fitted relaxation time scale is in good accord with the sub-second switching time visualized in Fig. 4J. This relaxation time scale is about two orders of magnitude longer than that in nanowires A–C, suggesting a larger $V_B$ value in this large nanowire.

**Extracting the potential barrier height for a mobile granular TLS.** Now consider the potential barrier height $V_B$. Recall that our measurements were carried out at 300 K. Thus, it is plausible to consider that the escape mechanism of a granular TLS is due to a thermal-activation process, with the relaxation rate given by

$$1/\tau_0 = f_0 \exp\left(-V_B/k_B T\right), \qquad (7)$$

where $f_0$ is the attempt frequency of the mobile nanocrystalline grain. Microscopically, the magnitude of $f_0$ is governed by the surrounding bonding potential profile and, in a solid, it is effectively characterized by the Debye temperature. The Debye temperature of ≈ 400 K in rutile $RuO_2$ gives $f_0 \approx 8\times10^{12}$ Hz (*47*). Then, from the inferred $\tau_0$ values and Eq. (7), we obtain $V_B \approx 0.6$ eV for the mobile nanograins in nanowires A–C, and $V_B \approx 0.7$ eV for the mobile nanograin in nanowire D. We should note that the extracted $V_B$ value is mainly governed by the exponential dependence in Eq. (7). For example, reducing the $f_0$ value by a factor of 2 will only lead to a small change of < 4% in the $V_B$ value. The reason for obtaining a (slightly) larger $V_B$ and hence a (much) longer $\tau_0$ in nanowire D, as compared to that in nanowires A–C, is as follows. The magnitude of $V_B$ for a granular TLS is essentially governed by those interfacial grain-boundary atoms which have the strongest bonding with neighboring lattice atoms. Due to a much larger grain size, the number of interfacial atoms in nanowire D far surpasses that in nanowires A–C. Thus, the probability for some interfacial atoms with relatively strong bonding is comparatively high, giving rise to a high $V_B$ barrier and long $\tau_0$ time. For comparison, we note that the potential barrier height between neighboring atoms in typical bulk metal crystals is on the order of $V_B \sim 10-30$ eV (*58*). Our low $V_B$ values inferred above clearly indicate relatively weak bonding strengths for the nanocrystalline grain boundaries in our $RuO_2$ nanowires. Indeed, our results are in good accord with the well-established reverse Hall-Petch effect, i.e., nanometer-sized granular systems are elastically soft and can be subject to easy plastic deformation (*6, 8, 9, 13, 14*). While $V_B$ is a crucial parameter which controls the mechanical strength of a nanograin system, to our knowledge no other existing experimental method can provide a reliable estimate of $V_B$ as done here. Moreover, in their studies of atomic TLSs in Cu nanobridges (*30*), Ralls and Buhrman previously inferred a $V_B$ value ($\simeq 66$ meV), one order of magnitude smaller than ours. This is expected, because an atomic TLS should be much easier to switch than a granular TLS. Thus, our observation of $V_B(\text{atomic TLS}) \ll V_B(\text{granular TLS}) \ll V_B(\text{bulk metal crystal})$ strongly suggests that our resistance fluctuations must be caused by moving clusters of atoms, i.e., mobile nanograins. Also, under the assumption of planar interfaces between grains, molecular dynamics simulation calculations of polycrystalline Ni films have revealed an effective activation energy of ~ 0.2 eV for the grain boundary sliding of 5 nm sized grains (*59*). While grain boundaries are curved in real samples (*60*), this theoretical value is supportive of our experimental results. Reversible grain boundary sliding, if any exists, may potentially be a source of granular TLSs.

Finally, although we observe one predominant mobile nanograin in each nanowire, it does not exclude the existence of (a few) more moving nanograins. As revealed in the HRTEM images shown in Figs. 2A and 2B (also fig. S1), several nanograins do exist in a nanowire and may be mobile simultaneously. However, owing to the measurement bandwidth of 0.03–1000 Hz employed in this study, only those mobile nanograins with relaxation time $\tau_0$ in the range 0.1 ms – 40 s can be detected. From Eq. (7), this range of $\tau_0$ corresponds to the potential barrier height in the range of $V_B \approx 0.5 - 0.9$ eV. We emphasize that our extracted $\tau_0$ and $V_B$ values are independent of any uncertainties that might occur in our estimate of the $N_{grain}$ value for a mobile nanograin.

**Discussion**

In summary, we have demonstrated an experimental method to study the atomic and, specifically, granular TLS dynamics in thin $RuO_2$ metal nanowires. We have observed that a nanocrystalline grain can spontaneously and reversibly switch between two or more metastable coordinate states. Both the temporal resistivity fluctuations and low-frequency noise power spectrum density were measured at 300 K. The atomic TLS density was extracted from the Hooge parameter $\gamma$ which characterized the magnitude of the 1/*f* noise. The size of a mobile nanograin, its relaxation time associated with a particular metastable configuration, and the bonding strengths of nanograin boundaries were obtained by analyzing the temporal resistivity fluctuations. Further improved estimates of the relevant electronic parameters together with proper modifications of Eqs. (4) and (5), where appropriate, for a specific material shall provide a fairly quantitative answer for the problem. Our high-resolution electrical method can be applied to a wide variety of nanodevices and 2D materials to sensitively infer the defect dynamics. We also remark that, with the downscaling of integrated circuits, the current densities in interconnects normally increase and electromigration often occurs, rendering a detrimental reliability problem (*61*, *62*). Our approach may be employed to investigate *real-time* mass diffusion at the atomic level. A good understanding of the migration mechanism will help the development of practical interconnect materials for the sub-10 nm integrated circuits.

## Materials and Methods

**Nanowire growth and structure characterizations.** Our $RuO_2$ nanowires A–C were grown on sapphire substrates using MOCVD with bis(ethylcyclopentadienyl)ruthenium as the source reagent, as described previously (*63, 64*). High purity oxygen was used as both carrier gas and reactive gas with the flow rate adjusted to 100 sccm. During deposition, the substrate temperature and pressure of the CVD chamber were controlled at 450°C and 10–50 torr, respectively. Nanowire D was grown via a thermal evaporation method (*65*), by applying Au nanoparticles (5–40 nm in diameter) as catalyst. Stoichiometric $RuO_2$ powder was placed in a quartz tube and heated to 920–960°C. An oxygen carrier gas, maintained at a pressure of 2 torr, transported the heated vapor to Si substrates held at 450–670°C. For HRTEM studies, nanowires were transferred to a copper grid, and HRTEM images were taken in an aberration-corrected STEM microscope (JEOL JEM-ARM200F with a Schottky gun) with an acceleration voltage of 200 kV.

**Electrical measurements.** For temporal resistance fluctuations and low-frequency noise measurements, individual $RuO_2$ nanowires were transferred to a 500 nm thick $SiO_2$ capped Si substrate. Cr/Au (10/100 nm) electrodes were deposited via standard electron-beam lithography to form contacts with single nanowires and connections to large Ti/Au (10/90 nm) pads which were pre-patterned on the Si substrate. The contact resistance between the nanowire and Cr/Au electrode was $\sim 100-500$ Ω. Thin copper wires then connected the Ti/Au pads (with silver paste) to the electrical pins (via soldering) on the sample holder of a standard cryostat insert.

Figure 2C shows a schematic for our measurement circuit. The circuit was modified from that utilized in our previous work (*11, 66*). Here we applied a 4-lead configuration, instead of a 5-lead bridge architecture. The measurement principle, employing the modulation and demodulation technique, was described by Scofield (*67*). The setup had the advantage that the $1/f$ noise contribution from the preamplifier (Stanford Research Systems Model SR560) was substantially minimized and the measurement sensitivity reached the lowest noise level of the preamplifier. For our sample resistance of ~ 0.3–10 kΩ, the noise contours of SR560 was flat in the frequency range $f \approx 1-100$ kHz. Thus, we applied an ac driving current with a carrier frequency $f_c \approx 3.1$ kHz. The sample resistance was $r(t) = \langle r \rangle + \delta r(t)$, where $\langle r \rangle$ being the averaged resistance, and $\delta r(t)$ being the temporal resistance fluctuations. Under an ac current $I(2\pi f_c t) \approx (V_0 / R_b)\sin(2\pi f_c t) \equiv I_0 \sin(2\pi f_c t)$, where $V_0 \sin(2\pi f_c t)$ was the output voltage from the lock-in amplifier (Stanford Model SR830), the voltage drop across the sample was $V_s = I_0 \sin(2\pi f_c t) \times [\langle r \rangle + \delta r(t)]$, with the second term $\delta V_s = I_0 \sin(2\pi f_c t) \times \delta r(t)$ being the voltage noise. Thus, the frequency components $f_i$ of $\delta r(t)$, after the Fourier transformation, was shifted (modulated) to $f_c \pm f_i$ (*67*), i.e., the frequency of $\delta V_s$ was centered at $f_c$. Note that $f_i \ll f_c$ for low frequency $\delta r(t)$. Our twisted, paired cryostat-insert wires had resistance $R \sim 50$ Ω and parasitic capacitance $C \sim 300$ pF to ground. For a sample resistance of 10 kΩ, the *RC* time constant ~ 3 μs rendered a bandwidth of ~ 50 kHz. Thus, the modulated $\delta V_s$ signal passed through the cryostat-insert wires with negligible attenuation. After $\delta V_s$ being amplified by SR560 with a gain $G_{\text{SR560}}$, the signal was sent back to SR830 for phase-sensitive detection. The SR830 multiplied the signal by the carrier frequency $f_c$, shifting (demodulating) the noise signal from $\sim f_c$ back to $f_i$ (*67*), and recovered the original frequency of $\delta r(t)$. (The time constant of SR830 was set at 100 μs, rendering a bandwidth of 1.6 kHz.) After demodulation, the "X component" signal was outputted to a dynamic signal analyzer (Stanford Model SR785) through the "CH1 OUTPUT" port with a gain $G_{\text{SR830}}$. The bandwidth of the X component output was 100 kHz. The SR785 read the voltage signal from the X component with a sampling rate set at 1024 Hz. This sampling rate ultimately governed the effective bandwidth of the measurements. The readings were stored in the buffers. A computer fetched the data from the buffers, dividing them by the total gain $G_{\text{SR560}} \times G_{\text{SR830}}$, and calculated the voltage noise PSD ($S_V$) using a LabView program. In each stage, the input noise was 4 nV Hz$^{-1/2}$ at 1 kHz for SR560 (the first stage), 6 nV Hz$^{-1/2}$ at 1 kHz for SR830 (the second stage), and $<10$ nV Hz$^{-1/2}$ for $f > 200$ Hz for SR785 (the third stage). The gain we employed $G_{\text{SR560}} = 100 - 1000$ resulted in an amplified input noise of 0.4–4 μV Hz$^{-1/2}$, which was much larger than the input noise of the second and the third stages. Thus, the noise level in the circuit was eventually limited by the input noise of SR560. In the setup, an isolated quiet ground was used and any possible ground loops were carefully eliminated. Before measuring nanowire samples, we checked the performance of our circuit by measuring the thermal noise of several ordinary metal-film resistors with various resistances. The measured $S_V$ conformed to the white noise $4k_B TR$ for $R > 1$ kΩ, whereas the measured noise level was limited by the input noise of SR560 for $R \leq 1$ kΩ. In addition, our measurements on metal-film resistors did not reveal any inherent oscillations or resistance jumps (see, for example, Figs. 4E and 4I), confirming that the resistance jumps shown in Fig. 4 must originate from $RuO_2$ nanowires but not from the experimental setup or electronics.

**Supplementary Materials**

Supplementary material for this article is available at ……

section S1. HRTEM images for additional $RuO_2$ nanowires.

section S2. Gaussian functions for resistivity histogram peaks.

section S3. Dwell time for a preferred resistivity value (state).

fig. S1. HRTEM images and SEAD patterns for $RuO_2$ nanowires.

fig. S2. Voltage noise PSD for metal film resistor.

fig. S3. Enlarged histogram plot depicting preferred resistivity states.

table S1. List of parameters for Gaussian functions.

**Acknowledgments**

**General**: The authors are grateful to the late Professor Ying-Sheng Huang for providing us the $RuO_2$ nanowires used in this study. We thank Lin-Lung Wei (L.L.W.) for taking the HRTEM images, and Yi-Chia Chou and L.L.W. for helpful discussion on the nanowire structure.

**Funding:** This work was supported by Taiwan Ministry of Science and Technology (MOST) through Grant numbers 103-2112-M-009-017-MY3 and 105-2923-M-009-005-MY2, and by the Taiwan Ministry of Education (MOE) ATU Plan.

**Author contributions:** S.S.Y. and J.J.L. conceived the experiment, W.Y.C. and S.S.Y. carried out the electrical measurements. All authors analyzed and discussed the data, and J.J.L. and S.S.Y. wrote the manuscript.

**Competing interests:** The authors declare that they have no competing interests.

**Data and materials availability:** All data needed to evaluate the conclusions in the paper are present in the paper and/or the Supplementary Materials. Additional data available from authors upon request.

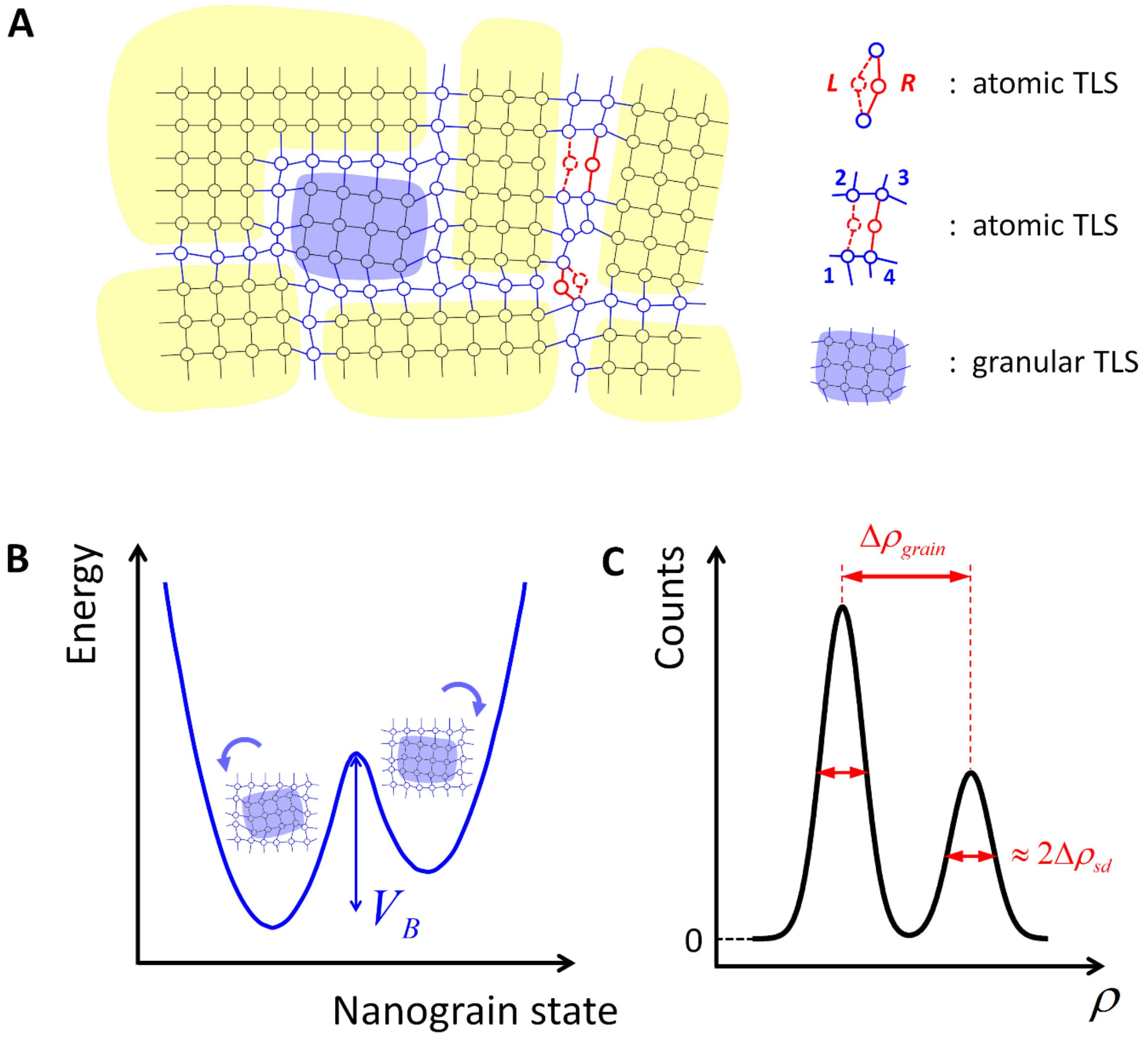

**Fig. 1. Schematic plots for dynamic defects and resistivity histogram.** (**A**) Schematic plot of atomic and granular two-level systems (TLSs). Circles denote atoms and short bars denote bonds. Black color denotes atoms and bonds inside a crystalline grain, and blue color denotes those at grain boundaries. Red circles depict atomic TLSs. Regions with shaded yellow background denote static crystalline grains, and the blue shaded background depicts a granular TLS. Top right: An atomic TLS can switch between the left (L) and right (R) bonding positions. An atomic TLS may also form bonding either with atoms 1 and 2, or with atoms 3 and 4. (**B**) Schematic plot of the energy diagram for a granular TLS with potential barrier height $V_B$ . (**C**) Schematic histogram plot for a set of temporal resistivity fluctuations $\rho(t)$ data. A sudden switch of a granular TLS causes a resistivity jump $\Delta\rho_{grain}$ . In each metastable nanograin configuration (coordinate state), the histogram peak is described by a Gaussian function with a standard deviation $\Delta\rho_{sd}$ .

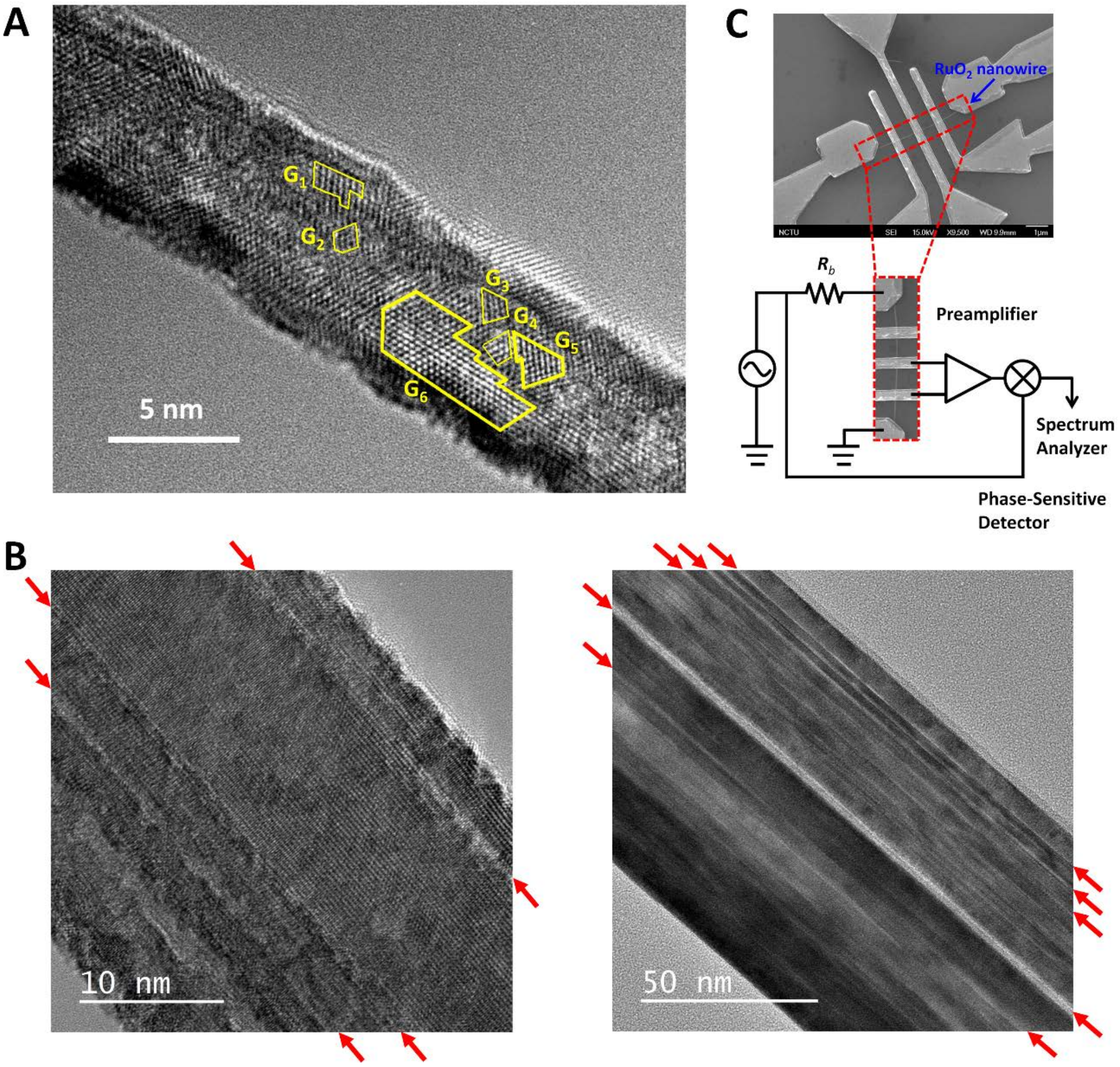

**Fig. 2. HRTEM and SEM images of $RuO_2$ nanowires and electrical measurement circuit.** (**A**) High-resolution TEM image of a $RuO_2$ nanowire from the same batch as nanowires A–C. Several nanocrystalline grains, labeled $G_1$–$G_5$, with diameter of ~ 1–3 nm are observed and indicated with yellow frames. $G_6$ denotes a comparatively large nanograin. (**B**) HRTEM images of a ~ 30 nm (left) and a ~ 90 nm (right) diameter $RuO_2$ nanowires from the same batch as nanowire D. Crystallites of tens nm in size are evident. The arrows indicate stacking faults. (**C**) (top) SEM image for nanowire A, and (bottom) a schematic for temporal resistance fluctuations and low-frequency noise measurement circuit. $R_b \approx 1\ \mathrm{M\Omega}$ is a ballast resistor.

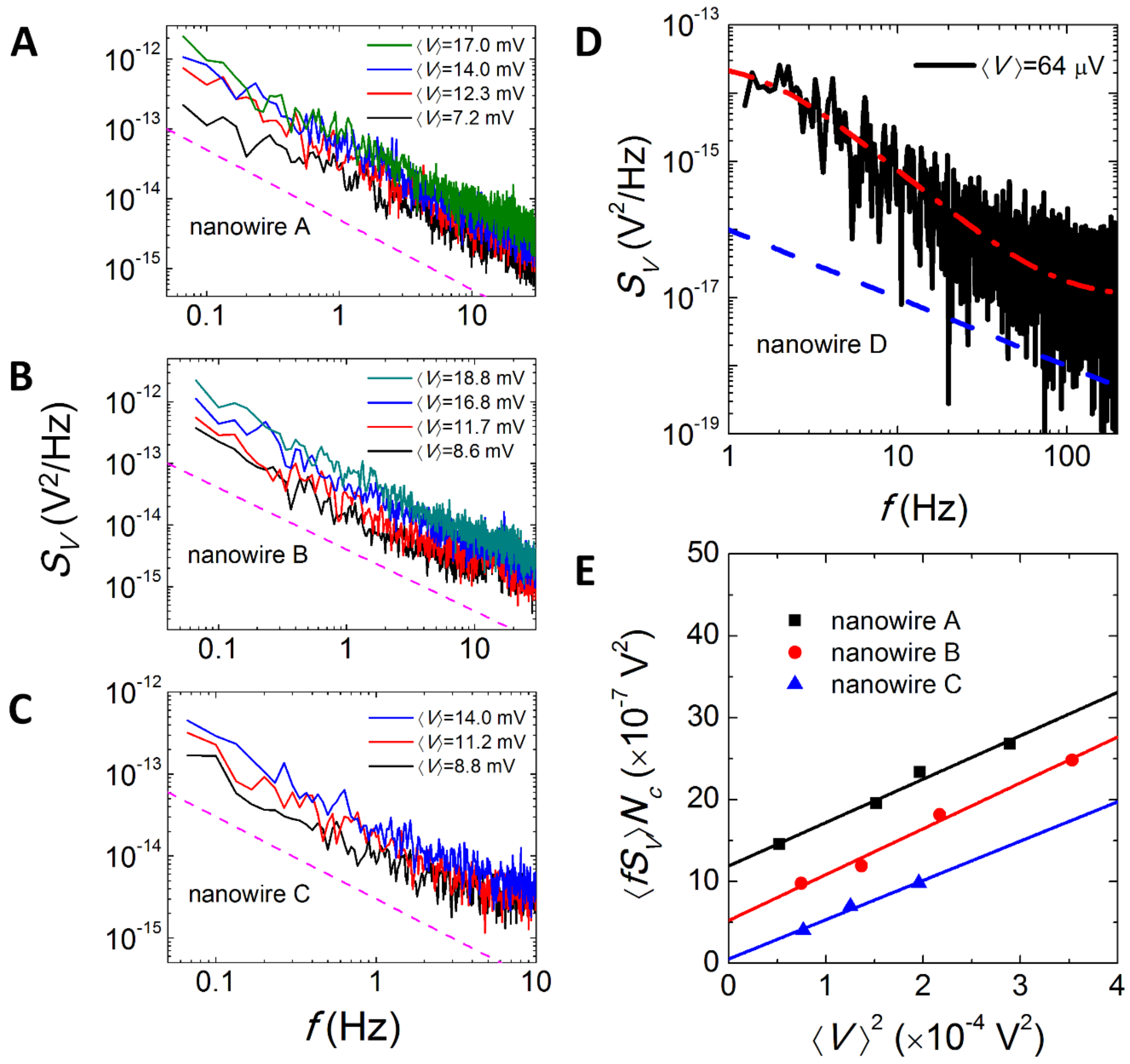

**Fig. 3. Power spectrum density and quadratic voltage dependence of $RuO_2$ nanowires.** (**A**)–(**D**) Power spectrum density $S_V$ for nanowires A–D measured at several bias voltages, as indicated. The straight dashed line in each panel indicates $1/f$ frequency dependence. While nanowires A–C obey a $1/f$ dependence, nanowire D obeys a Lorentzian dependence (red dashed-dotted curve) as predicted by Eq. (6), with $A = 3\times10^{-13}$ $V^2$, $\tau = 0.10$ s, and $S_V{}^0 = 1\times10^{-17}$ $V^2$/Hz. (**E**) Variation of $\langle fS_V\rangle N_c$ with $\langle V\rangle^2$ for nanowires A–C. $\langle fS_V\rangle N_c$ was averaged over 0.06 to 1 Hz. Data for nanowires A and B are offset for clarity. The straight lines are linear fits to Eq. (1).

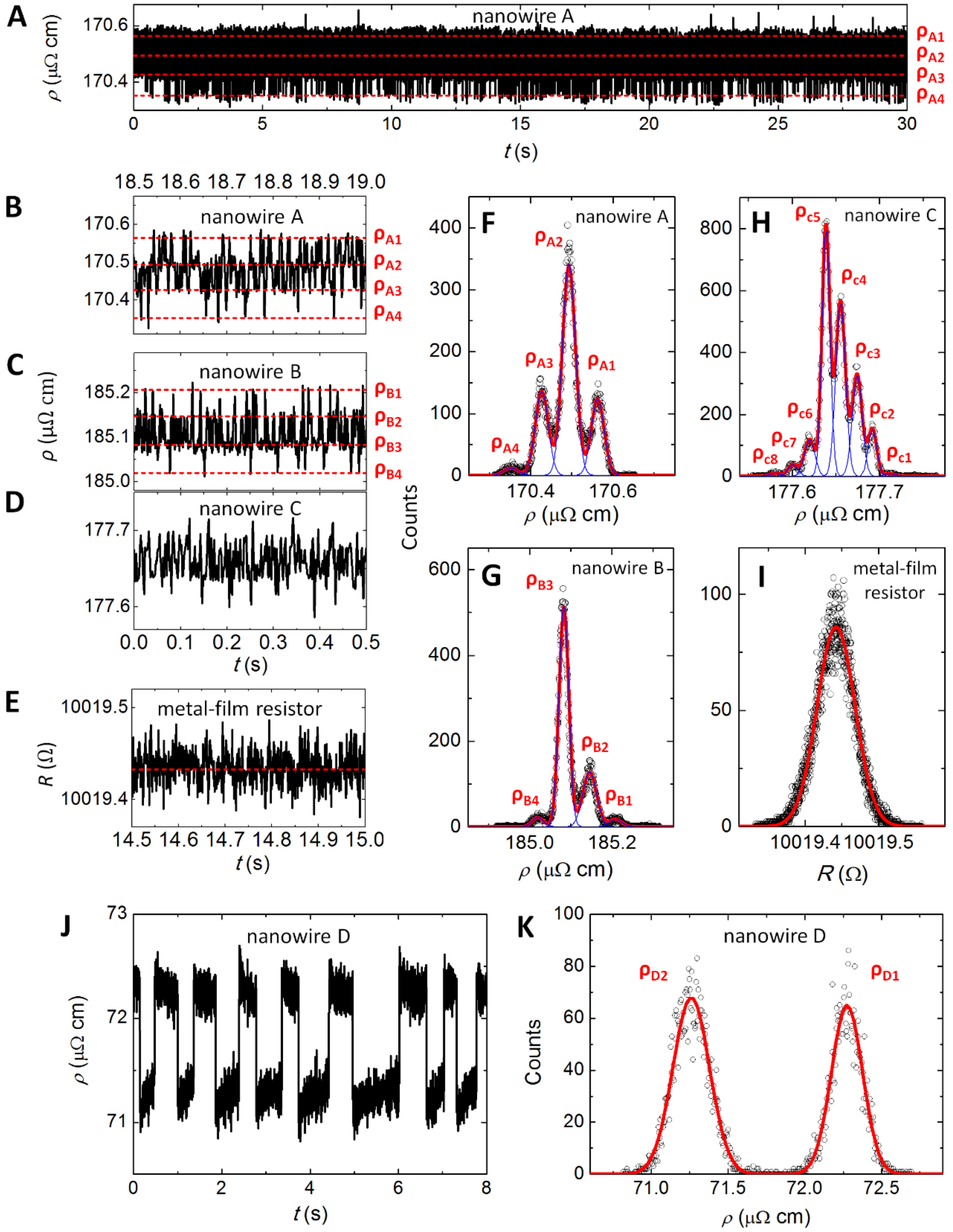

**Fig. 4. Temporal resistivity fluctuations and histogram plots.** (**A**) Temporal resistivity fluctuations for nanowire A recorded over 30 s. The four horizontal red dashed lines indicate the preferred resistivity values for the four histogram peaks potted in (**F**). (**B**)–(**E**) Temporal resistivity fluctuations for nanowires A–C and a 10 kΩ metal-film resistor recorded over 0.5 s, as indicated. Several preferred

resistivity states were observed in nanowires A–C, while only one state in the metal-film resistor. The dashed lines in (**B**) are the same as those in (**A**). The time period in (**B**) corresponds to 18.5–19.0 s in (**A**). (**F**)–(**I**) The corresponding histogram plots for nanowires A–C and the 10 kΩ metal-film resistor, as indicated. The blue curves in (**F**)–(**H**) are Gaussian functions and the red curve is the sum of the blue curves. The red curve in (**I**) is a Gaussian function. (**J**) Temporal resistivity fluctuations for nanowire D. Two preferred resistivity states are evident. (**K**) The corresponding histogram plot for nanowire D. The red curves are Gaussian functions. (Gaussian functions for all histogram peaks in (**F**)–(**I**) and (**K**) are listed in Supplementary Materials S2, table S1).

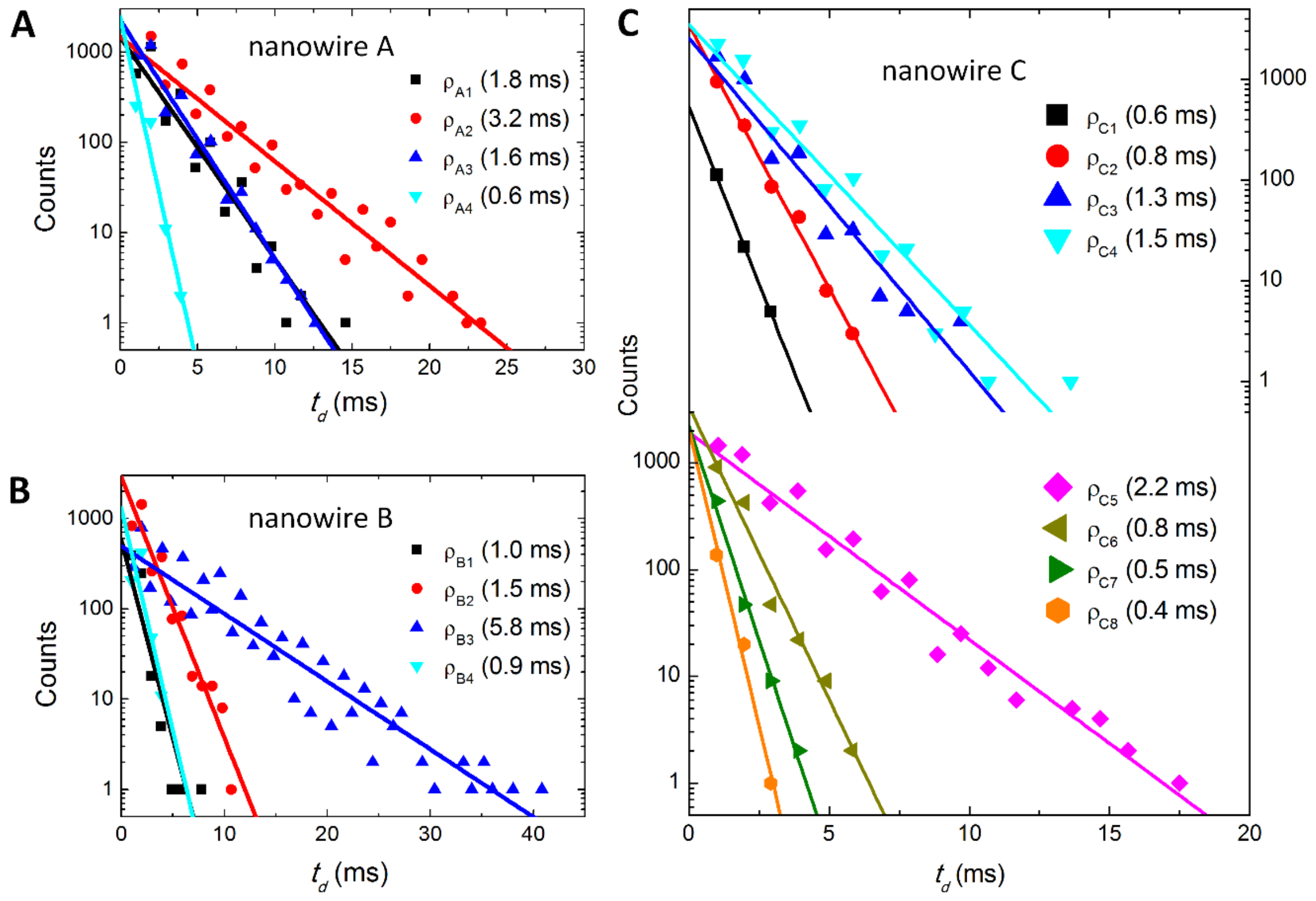

**Fig. 5. Histogram plots of dwell time.** Histogram plots of dwell time $t_d$ for each preferred resistivity state observed in nanowires A–C, as indicated. The straight lines are least-squares fits to the relaxation-time approximation $P(t) = P(0)\exp(-t/\tau_0)$, see text. The extracted $\tau_0$ value for each preferred resistivity state is listed in the parentheses in unit of ms in each panel.

**Table. 1. Relevant parameters for nanowires A–D.** $d$ is diameter, $L$ is the length between voltage probes, $\rho$ is resistivity, $\gamma$ is Hooge parameter, and $n_{TLS}$ is atomic two-level system density. Nanowires A–C were grown via MOCVD, while nanowire D was grown via thermal evaporation.

| Nanowire | $d$ (nm) | $L$ (μm) | $\rho$(300 K) (μΩ cm) | $\gamma$ | $n_{TLS}$ ($m^{-3}$) |
|---|---|---|---|---|---|
| A | 20 | 1.0 | 171 | $5.3\times10^{-3}$ | $2.9\times10^{26}$ |
| B | 20 | 1.1 | 185 | $5.6\times10^{-3}$ | $3.6\times10^{26}$ |
| C | 20 | 2.1 | 178 | $4.8\times10^{-3}$ | $2.9\times10^{26}$ |
| D | 90 | 3.8 | 71.7 | --- | --- |

# Supplementary Materials:

# Probing nanocrystalline grain dynamics in nanodevices

Sheng-Shiuan Yeh,[1] Wen-Yao Chang,[1] and Juhn-Jong Lin[1,2]

[1]Institute of Physics and [2]Department of Electrophysics, National Chiao Tung University, Hsinchu 30010, Taiwan

## S1. HRTEM images for additional $RuO_2$ nanowires

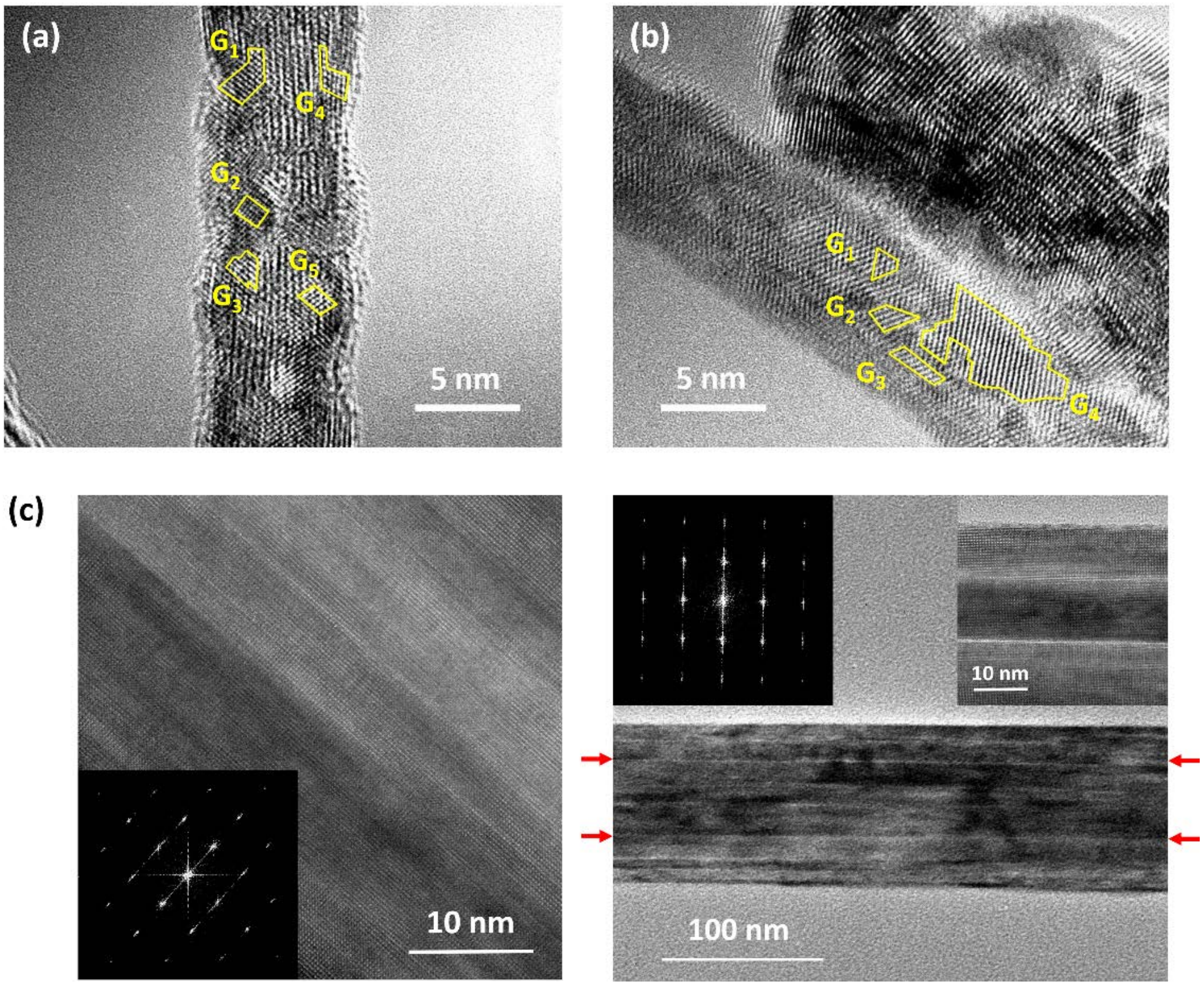

Figure S1

Apart from nanowires A–D discussed in the main text, we have studied additional $RuO_2$ nanowires. Figures 1(a) and 1(b) show HRTEM images measured on two different nanowires from the same batch as nanowires A–C. Nanocrystalline grains are observed. Figure S1(a) shows five nanograins (labeled $G_1$–$G_5$) of ~ 1–3 nm in size. Figure S1(b) shows four nanograins (labeled $G_1$–$G_4$) of ~ 1–7 nm in size. Figure S1(c) shows HRTEM images and selected-area electron diffraction (SAED) patterns measured on two nanowires from the same batch as nanowire D. The left image has been measured on the ~ 90 nm diameter nanowire illustrated in the right frame of Fig. 2B. The right image has been measured on another nanowire, with diameter ~ 100 nm. Stacking faults (indicated by arrows) are visible in both nanowires. The SAED patterns indicate that the structure is single crystalline.

**S2. Gaussian functions for resistivity histogram peaks**

In the Figs. 4(f)–(h), 4(k), and 4(i) in the main text, the histograms were plotted with interval widths of $8\times10^{-4}$, $7\times10^{-4}$, $8\times10^{-4}$, $4\times10^{-3}$ μΩ cm, and $2\times10^{-4}$ Ω, respectively. We find that each histogram peak can be described by a Gaussian function of the form $f_i(\rho)=a_i e^{-(\rho-\rho_i)^2/2c_i^2}$, where $i$ denotes the *i*th peak in the histogram plot, $a_i$ is a (non-normalized) prefactor characterizing the peak height, $c_i$ is the standard deviation characterizing the Gaussian rms width, and $\rho_i$ is the averaged resistivity of the *i*th peak. Our least-squares fitted values for these parameters are listed in Table S1. We obtain the fitted values $c_i=(1.68\pm0.12)\times10^{-2}$, $(1.42\pm0.25)\times10^{-2}$, $(5.53\pm0.96)\times10^{-3}$, and $(1.10\pm0.09)\times10^{-1}$ μΩ cm for nanowires A–D, respectively. That is, for a given nanowire, all the histogram peaks can be described by a set of Gaussian functions with standard deviations ($\Delta\rho_{sd}$) similar to within $\pm20\%$.

The Gaussian function (red curve) for the 10 kΩ metal-film resistor shown in the Fig. 4(i) in the main text is $f(R)=ae^{-(R-R_0)^2/2c^2}$, with $R_0=10019$ Ω, $a=86$, and $c=0.0257$ Ω. Figure S2 shows the noise PSD for this ordinary resistor. A 1/*f* dependence is evident in the frequency range 0.1–20 Hz.

**Table S1**

| Nanowire | A | B | C | D |
|---|---|---|---|---|
| $\rho_1$ (μΩ cm)<br>$a_1$, $c_1$ (μΩ cm) | 170.563<br>125, 0.0156 | 185.209<br>18, 0.0159 | 177.712<br>9, 0.00618 | 72.275<br>65, 0.101 |
| $\rho_2$ (μΩ cm)<br>$a_2$, $c_2$ (μΩ cm) | 170.493<br>341, 0.0159 | 185.146<br>128, 0.0167 | 177.692<br>153, 0.00488 | 71.26<br>68, 0.119 |
| $\rho_3$ (μΩ cm)<br>$a_3$, $c_3$ (μΩ cm) | 170.427<br>137, 0.0159 | 185.083<br>514, 0.0117 | 177.6745<br>329, 0.00563 | – |
| $\rho_4$ (μΩ cm)<br>$a_4$, $c_4$ (μΩ cm) | 170.352<br>12, 0.0179 | 185.02<br>22, 0.0144 | 177.6555<br>565, 0.00547 | – |
| $\rho_5$ (μΩ cm)<br>$a_5$, $c_5$ (μΩ cm) | – | – | 177.6388<br>809, 0.00457 | – |
| $\rho_6$ (μΩ cm)<br>$a_6$, $c_6$ (μΩ cm) | – | – | 177.62<br>119, 0.0059 | – |
| $\rho_7$ (μΩ cm)<br>$a_7$, $c_7$ (μΩ cm) | – | – | 177.601<br>38, 0.00564 | – |
| $\rho_8$ (μΩ cm)<br>$a_8$, $c_8$ (μΩ cm) | – | – | 177.5803<br>10, 0.00648 | – |

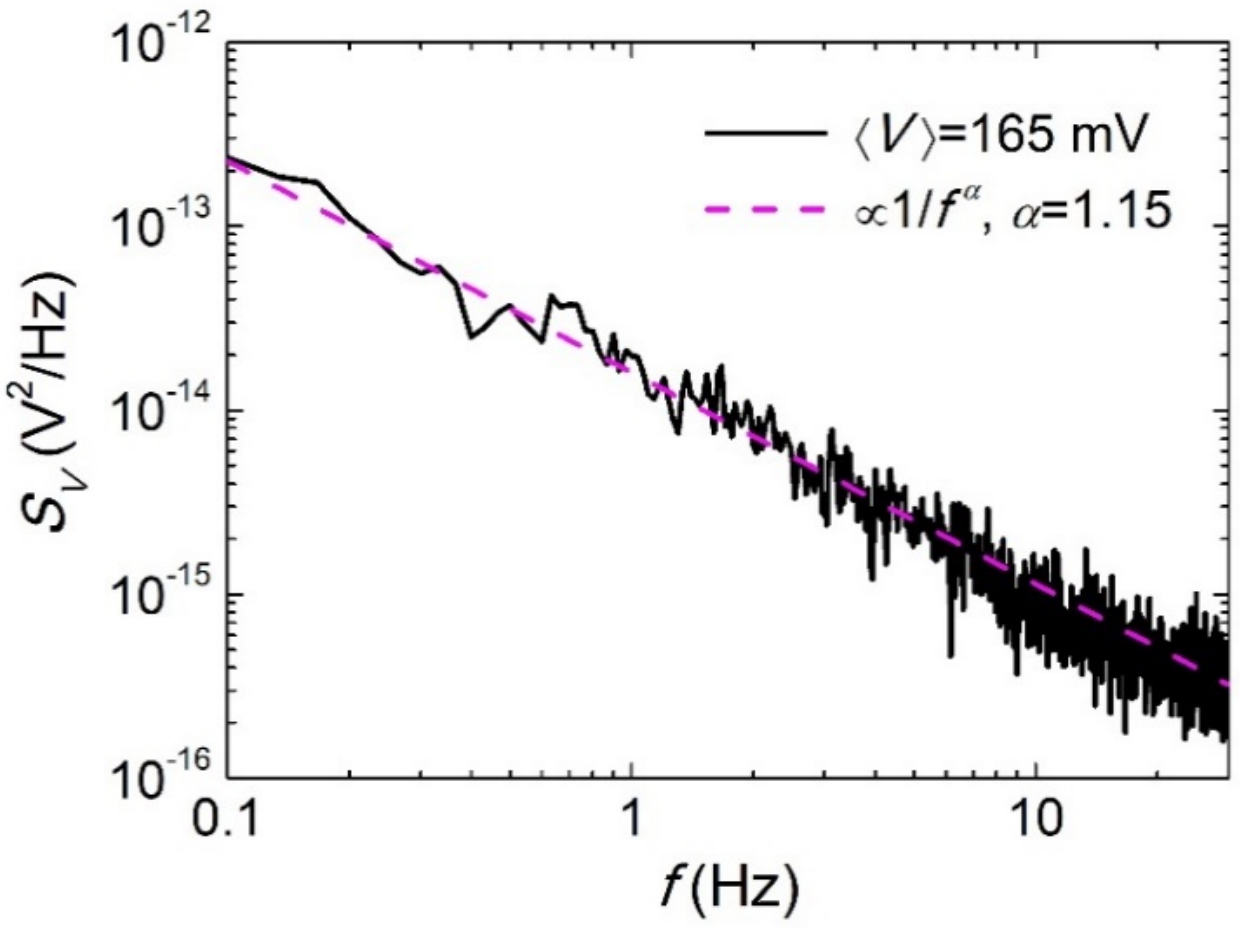

Figure S2

## S3. Dwell time for a preferred resistivity value (state)

The dwell time $t_d$ is the time period that a granular TLS spends in a certain preferred resistivity value (state). A preferred resistivity state is defined by the resistivity reading $\rho(t)$ of a nanowire falling in a given range of value. We choose this range of value to be the resistivity between the crossing points of the Gaussian functions which describe the neighboring histogram peaks. For example, in the case of nanowire A replotted in Fig. S3, the granular TLS is said to be in the $\rho_{A2}$ state when a $\rho(t)$ reading falls between $\rho_{A2,L}$ and $\rho_{A2,R}$.

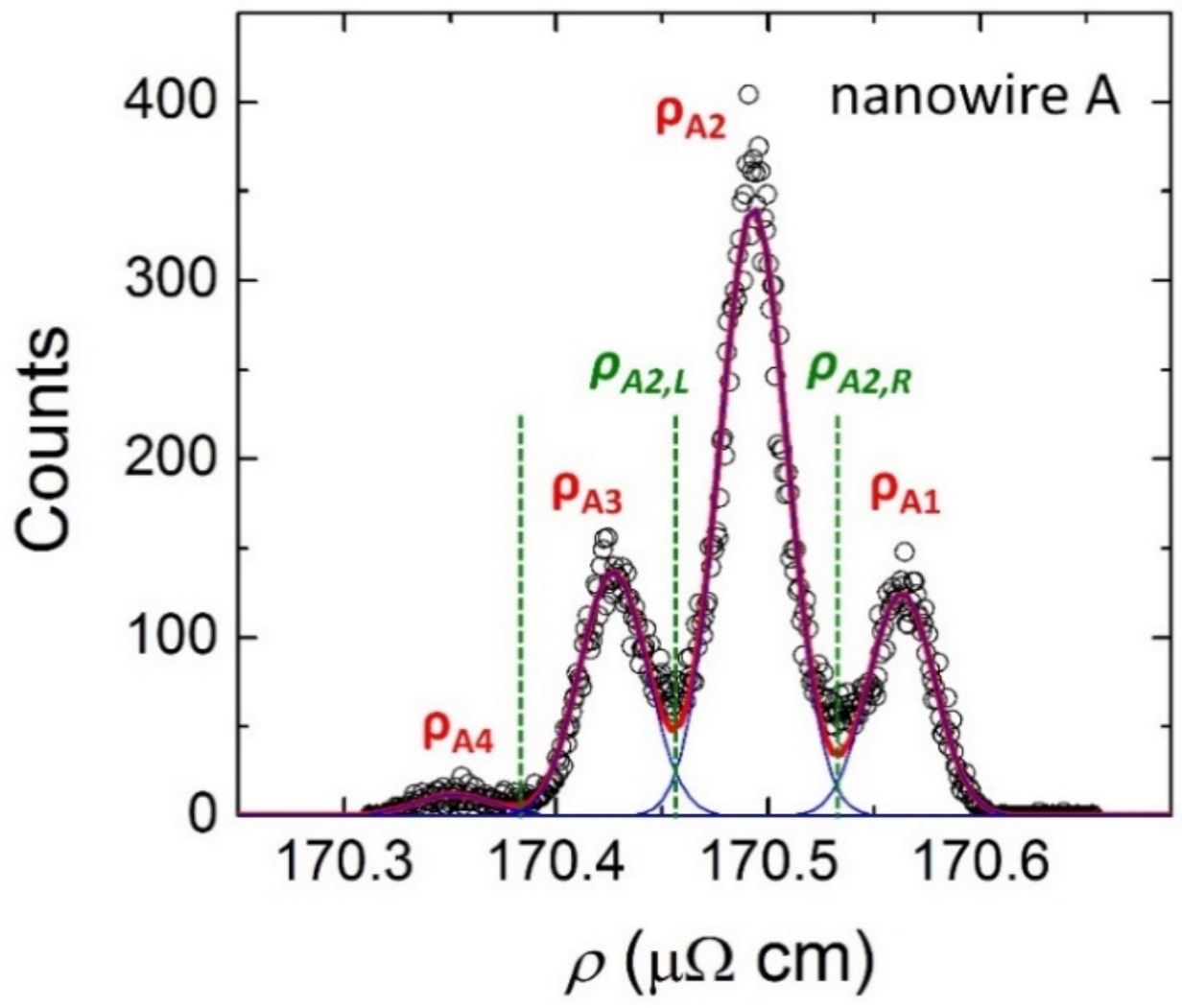

Figure S3

In our measurements of nanowires A–C, a $\rho(t)$ data set recoded over 30 s contains 30,720 readings, ensuring the occurrence of many “events” associated with each preferred resistivity state. For example, consider a particular event taking place in nanowire A, its dwell time for remaining in the $\rho_{A2}$ state is given by $t_d = t_2 - t_1$, where $t_1$ denotes the instant when a resistivity reading of $\rho_{A2,L} < \rho(t_1) < \rho_{A2,R}$ is first recorded (the beginning of an event), and $t_2$ denotes the instant when a resistivity reading of either $\rho(t_2) < \rho_{A2,L}$ or $\rho(t_2) > \rho_{A2,R}$ is subsequently recorded (the ending of the event). By this way, we have traced the $t_d$ value for every event in nanowires A–C and plotted the variation of event number (“counts”) with $t_d$ for each preferred resistivity state, as shown in the Fig. 5 in the main text.